\documentclass[twocolumn,aps,prb,showpacs,preprintnumbers,amsmath,amssymb]{revtex4-1}
\usepackage{natbib}
\usepackage{graphicx}
\usepackage{bm}
\usepackage{gensymb}
\usepackage{color}

\begin{document}

\title{Coexistence of long-range orders in a Bose-Holstein model} 

\author{Satyaki Kar} 
\author{Sudhakar Yarlagadda}

\affiliation{Saha Institute of Nuclear Physics, Salt Lake, Kolkata 700064, India.}
\date{\today}

\begin{abstract}
Exploring supersolidity in naturally occurring and artificially designed
systems has been and will continue to be an area of immense interest.
Here, we study how superfluid and charge-density-wave (CDW) states cooperate or compete
in a minimal  model for hard-core-bosons (HCBs) coupled
locally to optical phonons: a two-dimensional Bose-Holstein model.
 Our study is restricted to the parameter regimes
of strong HCB-phonon coupling and non-adiabaticity. We use
Quantum Monte Carlo  simulation (involving stochastic-series-expansion technique) 
to study 
 phase transitions and to investigate whether we have homogeneous or phase-separated coexistence.
The effective Hamiltonian involves, besides a nearest-neighbor hopping and a nearest-neighbor
repulsion,   sizeable double-hopping terms (obtained from second-order perturbation).
At densities not far from half-filling, in the parameter regime
where the double-hopping terms are non-negligible (negligible) compared to the nearest-neighbor hopping,
we get checkerboard-supersolidity (phase separation) with CDW being characterized by ordering
wavevector $\vec{Q}=(\pi,\pi)$.
\end{abstract}
\pacs{71.38.-k, 67.80.kb, 71.45.Lr, 74.20.Mn  } 
\date{\today}
\maketitle   

\section{Introduction}

Whether diagonal long range orders [such as charge-density-wave (CDW) and spin-density-wave (SDW)]
and off-diagonal long range orders [such as superconducting and superfluid (SF) states] 
can coexist homogeneously in correlated electronic systems is a central issue in condensed matter physics. 
Coexistence of superconductivity and CDW has been studied in 
 three-dimensional
systems \cite{blanton} (such as 
 ${\rm BaBiO_3}$ doped with ${\rm K}$ or ${\rm Pb}$), quasi-two-dimensional systems \cite{withers}
(such as the dichalcogenide ${\rm 2H-TaSe_2}$ and ${\rm NbSe_2}$)
and quasi-one-dimensional systems \cite{ong,abbamonte} (such as the trichalcogenide ${\rm NbSe_3}$ 
and doped spin ladder ${\rm Sr_{14}Cu_{24}O_{41}}$).

Supersolidity, defined as homogeneous coexistence of superfluidity and crystalline order,
was theoretically proposed more than 4 decades ago \cite{leggett} and has been a subject of debate since then. 
The first experimental claim \cite{kim} of observing supersolidity in helium-4 further intensified the debate
and enhanced the interest in understanding the phenomena. The general consensus \cite{chan,prokofev1}
is that a supersolid (SS)
is not realizable in a perfect hcp crystal of ${\rm {^4He}}$; 
nevertheless, superflow can occur
 along vacancies that
can collect near extended structural defects \cite{prokofev1,prokofev2}.
Thus, the occurrence of supersolidity in bulk solid helium-4 is being seriously questioned.

Cold-atom systems  offer another opportunity for realization of supersolidity.
Theoretically \cite{scalettar1,scalettar2,kedar1,kedar2,sheng,sengupta,sarma1}, 
lattice bosons with various types of interactions in diverse geometries 
have yielded
supersolidity. However, there has been no experimental creation of optical lattices with 
effective long-range interactions that produce supersolidity.
Furthermore, experimental techniques to detect signatures of supersolidity
also need to be developed for optical lattices \cite{sarma2}.  
 
 There have been numerous studies of supersolidity involving hard-core-bosons (HCBs)
 \cite{scalettar1,scalettar2,kedar1,kedar2}.
A lattice model for quantum liquids, such as the interacting bosonic helium-4 at low temperatures,
 needs a hard-core constraint
  to account for the exclusion of occupation of more than one atom at each lattice point \cite{matsubara,matsuda}.
  The behavior of the ground state and low temperature excitations of such systems
 are largely controlled by couplings with phonons.
Furthermore,   local Cooper pairs [comprising of two electrons (of opposite spin) at a site]
can also be regarded as HCBs \cite{ng}. In Bismuthates,  such HCBs couple to the breathing mode of the
oxygen cage surrounding the Bismuth ions  \cite{varma,tvr}.

Here, in this article, we study a two-dimensional (2D) Bose-Holstein (BH) model for HCBs on a square lattice where
 they can hop to nearest-neighbor (NN) sites 
and experience the HCB-phonon interactions via a Holstein-type term. Previously,  exact diagonalization
 calculations were done on this model \cite{sanjoy} for a small system (i.e., $4\times4$ lattice) to study  
the resulting phase diagram. Here we use stochastic-series-expansion (SSE) based  quantum Monte Carlo (QMC) technique
 to simulate  large-size lattices so that various phases in the thermodynamic limit 
can be identified more clearly.
Unlike the $t-V$ model,  a SS
 is realized in our BH system due to non-negligible transport within the same sublattice.
At densities not far from half-filling and at sufficiently large HCB-phonon couplings, phase coexistence
occurs; furthermore, in the phase-coexistence region, the system tends to phase separate at stronger couplings.

Our paper is organized as follows: section II deals with a  discussion of
the BH Hamiltonian, its transformations and its mapping to the equivalent spin model.
 Section III covers a description of the  numerical method and the observables employed to characterize the orderings. 
In section IV, we detail our results and the corresponding analysis, both with 
and without the presence of same-sublattice hopping terms.
Lastly, in section V, we summarize our results and draw conclusions.

\section{Formulation}

The BH Hamiltonian is given by
\begin{eqnarray}
\!\!\!\!\!\!\!\! H=-t\sum_{j,\delta}b_j^\dagger b_{j+\delta}+\omega_0\sum_j a_j^\dagger a_j +g\omega_0\sum_j n_j(a_j+a_j^\dagger) ,
\end{eqnarray}
where $a_i$ and $b_i$ denote the annihilation operators for phonons and HCB particles, respectively, and $n_i$ $(\equiv b_i^\dagger b_i)$ is the number operator
for HCBs at site ${i}$. Furthermore, $t$ is the amplitude for hopping at NN sites denoted by $\delta$ and $\omega_0$ is the frequency of optical phonons \cite{mahan}.
An effective Hamiltonian for the HCB particles is obtained by first transforming this BH Hamiltonian to the polaronic frame of reference
(using the Lang-Firsov transformation) and then performing perturbation theory as detailed in Refs. \onlinecite{sanjoy,sahinur}.
Interestingly, second-order perturbation theory yields a two-step hopping which produces{, in a 2D square lattice,} both next-nearest-neighbor (NNN) and next-to-next-nearest neighbor (NNNN) hopping terms besides the usual NN hopping term in the Hamiltonian. Moreover, a NN repulsion also results from two-step virtual hopping back and forth. 

So we get an effective $t_1-t_2-t_3-V$ Hamiltonian for HCB particles on a 2D square lattice  \cite{sanjoy}:
\begin{align}
H_e&=-g^2\omega_0\sum_jn_j-t_1\sum_{j,\delta}b_j^\dagger b_{j+\delta}-t_2\sum_{j,\delta'}b_{j}^\dagger b_{j+\delta'}\nonumber\\&
-t_3\sum_{j,\delta''}b_{j}^\dagger b_{j+\delta''}-\frac{V}{2}\sum_{j,\delta}n_j(1-n_{j+\delta}) ,
\label{He}
\end{align}
where $\delta'$ and $\delta''$ denote NNN and NNNN sites respectively; 
$t_1=t$exp$(-g^2)$, $t_2=(2t_1^2/\omega_0)f_1(g)$, $t_3=t_2/2$ and $V=(t_1^2/\omega_0)[4f_1(g)+2f_2(g)]$; 
$f_1(g)\equiv\sum_1^\infty g^{2n}/(n!n)$  and 
$f_2(g)\equiv\sum_{n,m=1}^\infty g^{2(n+m)}/[n!m!(n+m)]$.
In the regime $g > 1$, we can make the approximations 
 $f_1(g) \sim \frac{e^{g^2}}{g^2}$ and $[f_2(g) +2f_1(g)] \sim \frac{e^{2g^2}}{2g^2}$ with the approximations 
becoming exact for $g \rightarrow \infty$. 
The small parameter is $t/(g\omega_0)$ and is obtained from $[V/\omega_0]^{1/2}$
(see Ref. \onlinecite{pankaj} for details);
our perturbation analysis is done in the nonadiabatic regime ( $t \le \omega_0$) and at strong coupling ($g > 1$).

Since all the hoppings are non-frustrated, 
a QMC simulation of the system does not suffer from the negative sign problem. We employ SSE technique in our QMC simulation 
 and investigate the co-existence or competition of CDW and superfluidity  in various  regimes of the parameter space.
Here we should mention that our $t_1-t_2-t_3-V$ model for HCBs is equivalent to an extended XXZ spin-1/2 Hamiltonian as shown below:
\begin{align}
H=&\sum_{<i,j>}[J_{1z}S_i^zS_j^z+\frac{J_{1xy}}{2}(S_i^+S_j^-+ {\rm H.c.})]+
\nonumber\\&\frac{J_{2xy}}{2}\sum_{<<i,j>>}(S_i^+S_j^-+{\rm H.c.})+\nonumber\\
&\frac{J_{3xy}}{2}\sum_{<<<i,j>>>}(S_i^+S_j^-+{\rm H.c.})-h_0\sum_iS_i^z ,
\label{xxz_ext}
\end{align}
where $<i,j>$, $<<i,j>>$, and $<<<i,j>>>$ stand for NN, NNN, and NNNN pairs respectively.
Furthermore, the operators for the HCBs are related to those of spin-1/2 particles as:
 $S_j^z=n_j-\frac{1}{2}$ and $S_j^+=b_j^\dagger$. A comparison of the parameters in Eqs. (\ref{He}) and (\ref{xxz_ext}) yields:
 $J_{1z}=V$, $J_{1xy}=-2t_1$, $J_{2xy}=-2t_2$, $J_{3xy}=-2t_3$ and $h_0=g^2\omega_0$.
Now, the magnetization of the system can be tuned by using an external magnetic field; then,
a term $-h J_{1xy} \sum_i S_i^z$ should be added to the Hamiltonian in Eq. (\ref{xxz_ext}) where the magnetic field $h$
is given in units of $J_{1xy}$. 

\begin{table}
\begin{center}
\caption{$J_{1z}$ and $J_{2xy}$ in terms of $J_{1xy}$ [in Eq. (\ref{xxz_ext})]
at various values of the HCB-phonon coupling $g$ and at $t/\omega_0=1$.}
\begin{tabular}  {|p{30pt}| p{30.00pt}| p{30.00pt}|p{30.00pt}|p{30.00pt}|p{30.00pt}|p{30.00pt}|p{30.00pt}|p{30.00pt}|p{30.00pt}|}     \hline
  ${\bf g}$ & 0.5 & 1.0 & 1.5 & 2.0 & 2.5 & 3.0 
  \\ \hline\hline ${\bf J_{1z}}$ & 0.444 & 1.355 & 2.725 & 8.017 & 45.485 & 478.571\\ \hline  ${\bf J_{2xy}}$& 0.415  &  0.970 &  0.960 &  0.647  &  0.395  &  0.255 \\ \hline
\end{tabular} 
\label{tab1}
\end{center}
\end{table}

Presence of hopping terms for HCBs indicates that superfluidity (i.e., spontaneous breaking of the global U(1) gauge symmetry) can exist in the system.
 On the other hand, a large interaction strength suggests the possibility of a CDW. 
Thus, our objective is to study the compatibility of these two long range orders.
Now, these two orders can coexist either in a phase separated form or homogeneously as a SS. 
It should be pointed out that, a $t-V$ model on a square lattice does not show a thermodynamically stable SS phase for HCBs \cite{scalettar1}.
On the other hand, striped SS behavior is found away from half filling when NNN repulsion ($V_2$) is considered in the
 $t-V_1-V_2$ model\cite{scalettar1,scal3,sengupta}.

\section{Numerical 
calculations  using SSE-QMC}

We now give details of the SSE-QMC simulation of our $t_1-t_2-t_3-V$ model, or equivalently, our extended XXZ spin-1/2 model.
Finding the phase diagram in the present problem requires exploring various limits of the parameters
(including high anisotropy in our spin model). In our numerical computations, we used directed loop update
for efficient sampling of the configurations \cite{sse1,sse2}. The ground state properties are captured by simulating
 at low enough temperatures, i.e., $\beta\sim L$ with $L$ being the linear dimension of the square lattice\cite{scal2}.
We employ $\beta=3L/2$ since our calculations with $\beta=2L$ yield the same values for the observables (within the error bars of the calculations).
From calculations involving various large system sizes,
we 
infer the results in the thermodynamic limit.

{
{ As can be seen from the expressions of the two-spin matrix elements for Heisenberg spin models used in various SSE-QMC studies \cite{sse1,sse2},
 a positive parameter $\epsilon$ is introduced 
to ensure the positivity of all the matrix elements (see Appendix A for details). 
This is necessary so that they can be treated as probabilities.
 The value of $\epsilon$ can also affect the autocorrelation time of the desired variable. We found that keeping the
numerical value of $\epsilon$  equal to at least a quarter of the anisotropy parameter $J_{1z}/J_{1xy}$, particularly near the transition region, 
 helps keep the data in various bins uncorrelated  when each bin contains data from large
 number of Monte Carlo sweeps (i.e., at least 15,00,000); for the value of $\epsilon$ used in the XXZ model, see Ref. \onlinecite{sse1}.}}

In this work we are concerned with the diagonal order parameter $S(\pi,\pi)$ [i.e, the structure factor at the Neel ordering vector $\vec{Q}=(\pi,\pi)$]
 and the off-diagonal order parameter of the SF density $\rho_S$. A general expression for $S(\vec{Q})$ (for our HCB system) is given as 
{
\begin{eqnarray}
S(\vec{Q})=\frac{1}{N}\sum_{i,j}e^{i\vec{Q}.(\vec{R}_i-\vec{R}_j)}(\langle n_i n_j \rangle - \langle n_i \rangle \langle n_j \rangle )  ,
\end{eqnarray}
}
 where 
 $\langle \rangle $ denotes the ensemble average.
Since $n_i$ (or $S_i^z$ in our spin model) are diagonal in the basis, a QMC average can be computed easily.

The SF density \cite{sandvik} is given by $\rho_S=1/N\left ( \partial^2F/\partial \theta^2 \right )_{\theta=0}$, where $F$ is the free energy in the presence 
of twisted boundary conditions with angle of twist $\theta$.
This is an off-diagonal order parameter. In a QMC calculation, the SF density along $x$-direction is calculated using $\rho_{Sx}
=<(N_x^+-N_x^-)^2>/{\beta N}$ 
where $N_x^+$ and $N_x^-$ represent the total no. of Hamiltonian operators 
transporting spin in the positive and negative $x$-directions, respectively \cite{sandvik}.

{A few benchmarking comparisons of our SSE results with exact results for a $J_1-J_2$ model are shown in the supplementary material.}

\section{Analysis of 
 results and technicalities} 
\begin{figure}[t]
\vskip .1 in
\begin{center}
\includegraphics[width=3.0in,height=2.0in]
{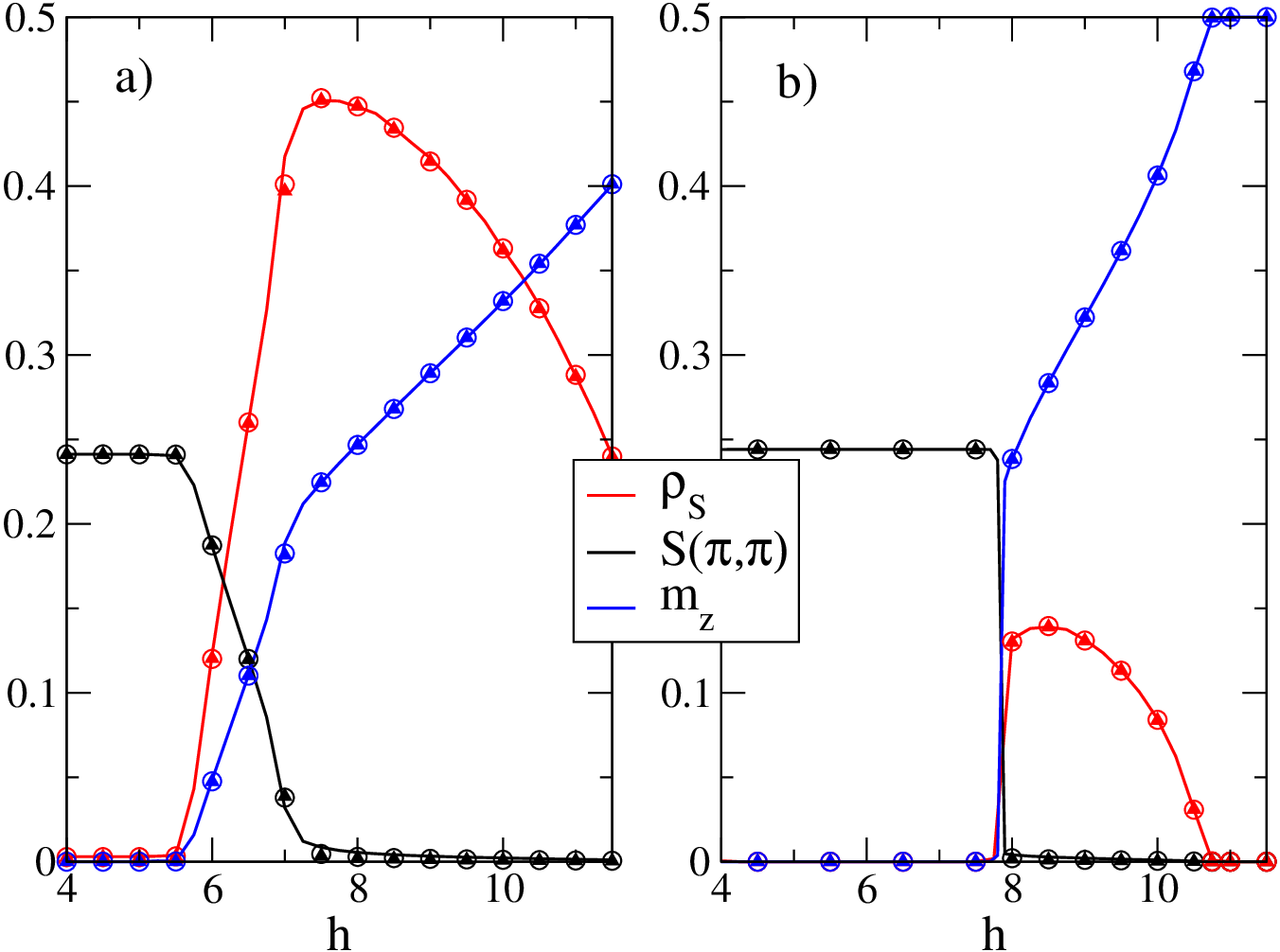}
\end{center}
\caption{ (Color online) {\bf Identifying sufficiently large system sizes to capture
quantum phase transition.} $\rho_S$, $S(\pi,\pi)$ and $m_z$ vs magnetic field h in $L\times L$ square
 lattices with L = 10 (line), 14 (circle) and 16 (triangle up) for $t/\omega_0 =1$ and
 $g=1.75$ when (a) $J_{2xy}=2J_{3xy}\neq 0$ 
 and (b) $J_{2xy}=J_{3xy}=0$.}
\label{Figure 1}
\end{figure}

Our calculations for the 2D $t_1-t_2-t_3-V$ model  
aim at obtaining the ground state phase diagram 
of the system and also at extending (to the thermodynamic limit) the findings presented by a Lanczos study on a small cluster in 
Ref. \onlinecite{sanjoy}. { We consider $t/\omega_0=1$ in the present study.}

To obtain the phase diagram for our system, the interplay between the diagonal and off-diagonal orders and
 the transition between them needs thorough investigation. As antiferromagnetic order breaks SU(2) symmetry whereas a 
SF phase breaks U(1) symmetry, a phase transition between them (according to Landau theory)
 cannot be of second-order type. Furthermore, a transition to a phase separated state occurs when
the system undergoes a first-order transition. 

\begin{figure}[t]
\begin{center}
\includegraphics[width=3.0in,height=4in]{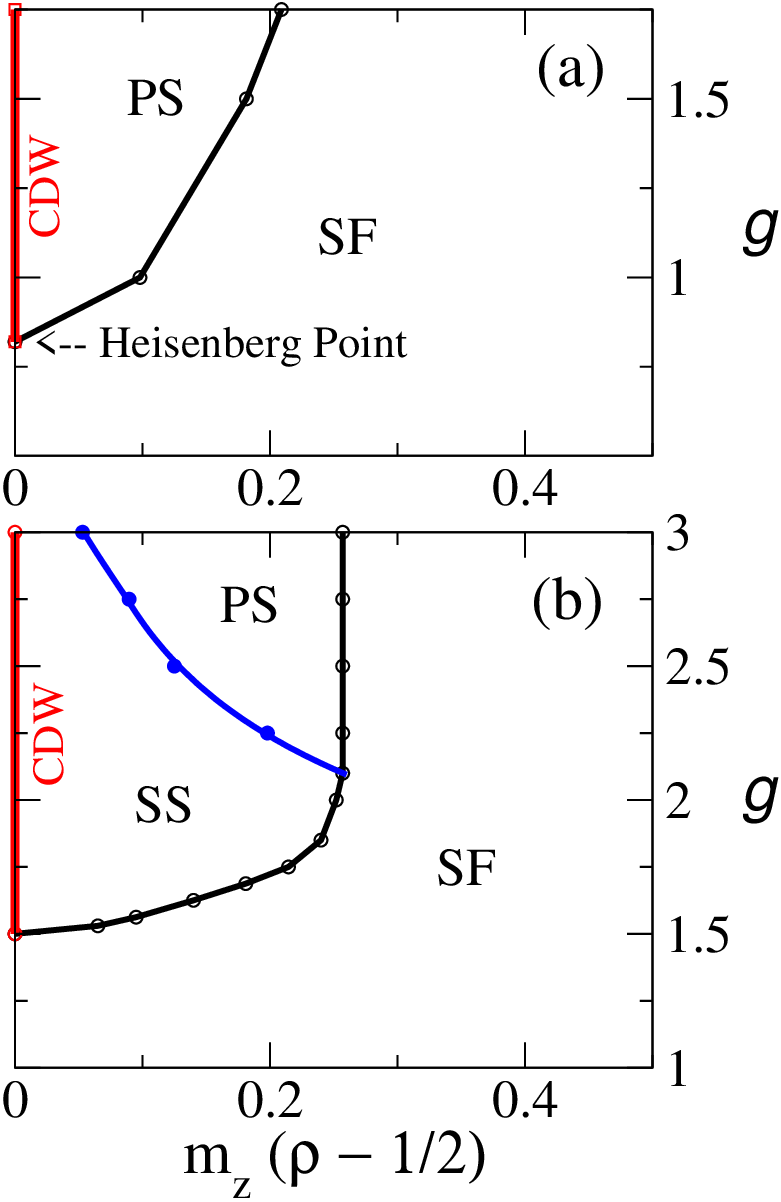}
\end{center}
\caption{(Color online) {\bf  Quantum phase diagrams at various magnetizations  (fillings)
 and adiabaticity ${\mathbf {t/\omega_0 = 1.0}} $}. The calculations are for a $16\times 16$ lattice
and for our BH system [using Eq. (\ref{He})] by (a) considering $t_2=t_3=0$ and  (b) including all
the interactions and hoppings.}
\label{Figure 2}
\end{figure}

\begin{figure}[b]
\vskip .1 in
\begin{center}
\includegraphics[
width=3.0in,height=1.8in]
{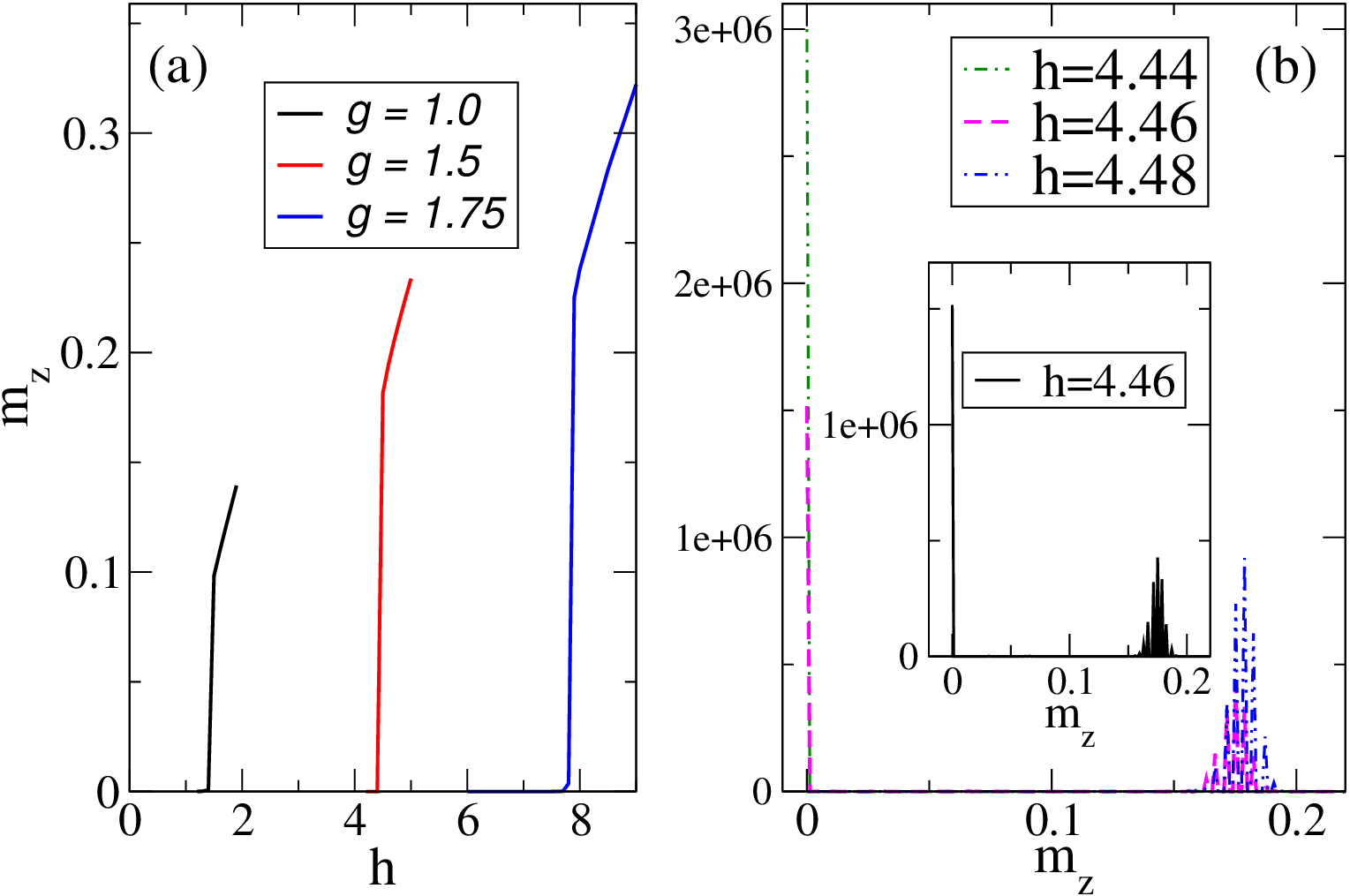}
\end{center}
\caption{(Color online) {\bf Magnetization plots showing discontinuous transitions in the XXZ model.}
 (a) Magnetization $m_z$ vs magnetic field $h$ in a $16\times16$ lattice at $\beta=1.5L$,
at $t/\omega_0 =1$, and for different values of $g$. 
(b) Magnetization histograms in the transition region at $g=1.5$; two peak
 structure of the histogram for $h=4.46$ is highlighted in the inset. }
\label{Figure 3}
\end{figure}

We work in the grand canonical ensemble where  there is no constraint on the HCB particle
 number (or the magnetization in the equivalent extended XXZ model).
In Fig.~\ref{Figure 1}, we show  the variation  of $m_z$, $S(\pi,\pi)$, and $\rho_{S}$  with magnetic field $h$ (expressed in units of $J_{ 1xy}$)
for g=1.75 for both our extended model
 and its XXZ version (i.e., $J_{2xy}=J_{3xy}=0$) in $L\times L$ square lattices with $L=10,~14$ and $16$.
 The comparison shows that the results for different size lattices almost coincide. 
At small values of the magnetic field $h$, 
the system  manifests  half-filling in the case of HCBs (or zero magnetization in the case of the equivalent spin model);
 the CDW phase is formed with maximum values of $S(\pi,\pi)$ with SF density simultaneously assuming
zero value. At large values of $h$, before the system is at complete filling, $S(\pi,\pi)$ decreases to zero while $\rho_S$ becomes finite 
manifesting the SF phase. 
In the intermediate magnetic field region, phase transition occurs with both the orders coexisting in our extended XXZ model.
 The plot of magnetization $m_z$ [or ($\rho -1/2$) where $\rho$ is the density]
in Fig.~\ref{Figure 1}(b), indicates a discrete jump during the transition in the case of the XXZ model.
 This is a typical signature for a first-order transition. Contrastingly, a continuous smooth increase in $m_z$ is observed for the extended XXZ model
clearly ruling out the possibility of a phase separation. 
Calculations using a canonical ensemble shows an inhomogeneous phase coexistence due to phase separation (PS) \cite{sanjoy}.
 
In the region adjacent to $m_z=0$ in Fig. \ref{Figure 1}(a), where overlap of $S(\pi,\pi)$ and $\rho_S$ are observed, the system displays a SS phase.
 Analysis of various large system sizes confirms the picture in Fig. \ref{Figure 1}(a)
that there is indeed a homogeneous coexistence of 
 CDW and SF long-range orders in the system.

Next, we will do a further detailed case-by-case study for the system with and without the effects of same-sublattice (i.e., NNN and NNNN) hoppings.

\subsection{
 Considering only NN hopping ($t_2=t_3=0$)}
Here we study the  case of $t_2=t_3=0$ in Eq. (\ref{He}), $i.e.$, the bare XXZ model. 
Calculations are done on a $16\times16$ lattice. 
We find that the system loses its CDW order at half-filling for values of the coupling $g$ below the critical value of $g_c=0.82$
 corresponding to the Heisenberg point of the XXZ model.
 For smaller values of $g$, superfluidity develops for all values of filling between 0 and 1.
The  phase diagram for the XXZ model is depicted in Fig.~\ref{Figure 2}(a).
Our calculated phase diagram of the XXZ model is compatible with Fig. 1, Fig. 2a and Fig. 2b of Ref. \onlinecite{schmid}.
At half filling (i.e., $\rho=1/2$ or $m_z=0$), the Heisenberg point denotes the boundary between CDW and SF phases.

Fig. \ref{Figure 3}(a) shows the jumps in magnetization $m_z$ as soon as we increase $g$ beyond $g_c$. 
 The magnitude of the jump increases as the coupling $g$ increases. This jump implies a first-order transition and
indicates a phase separated coexistence of the CDW and SF phases. A histogram analysis can also capture the jump in $m_z$ values.
 In Fig. \ref{Figure 3}(b), plotted for $g=1.50$,  the $m_z$ histograms change when  magnetic field values are varied.
For instance, when the magnetic field is set at $h=4.44$, $m_z=0$ even when  large number of Monte Carlo sweeps were used in our QMC computation.
On increasing the magnetic field to $h=4.46$, the $m_z$ values start showing a double-peaked structure (with one peak at $m_z=0$ and
another peak  at $m_z \approx 0.18$)
which is indicative of phase separation [see also inset in Fig. \ref{Figure 3}(b)]. 
Then, at a slightly higher magnetic field of $h=4.48$,  all
 the $m_z$ values seem to be centered around a mean value close to 0.18. 
This manifests the first-order transition. 
All the discontinuous transitions are due to  inhomogeneous coexistence of the CDW and SF phases.

\subsection{
Considering all  hoppings}

The phase diagram for our BH model [obtained from Eq. (\ref{He})] is depicted
in Fig.~\ref{Figure 2}(b).
On including the effects of NNN and NNNN hoppings (i.e., $t_2= 2 t_3 \neq 0$)
in the $t_1-t_2-t_3-V$ model of Eq. (\ref{He}), there can be a difference in the densities in the two sublattices
owing to NN repulsion and 
same-sublattice hopping. 
 As shown in Table. \ref{tab1}, at intermediate values
of $g$ (i.e., $g \sim 1$ ), NNN hopping is comparable to NN hopping; consequently,  a SS state can occur.
 Whereas at larger values of $g$ (i.e., for $g > 2.5$), NNN hopping
is fairly smaller than NN hopping and we can expect the same behavior as in the $t-V$ (or the XXZ) model.
On account of particle-hole symmetry in our model, 
the phase diagram is symmetric about half-filling.

We will now examine various features of the phase diagram of Fig. \ref{Figure 2}(b).
The phase diagram for our  model was obtained by identifying the transition regions and the nature of the various phases.
The variation of magnetization with magnetic field in a $16\times16$ 
lattice is shown in Fig.~\ref{Figure 6} in the transition region. The results for $g=1.5, ~1.75,~\&~2.0$ 
show that the magnetization increases gradually without any jump as the magnetic field is increased.
Hence, in Fig.~\ref{Figure 6}(a), for $g=1.75~\&~2.0$  CDW and SF phases coexist homogeneously 
resulting in a SS state. Here we must mention that Fig. \ref{Figure 1}(a) (depicting simultaneous coexistence
of CDW and SF phases through non-zero values of $S(\pi,\pi)$ and $\rho_S$) corroborates this conclusion.

As a matter of fact, large anisotropy [i.e., large values of $V/t_1$ in Eq. (\ref{He})] in the model requires large simulation time. 
Thus, as we increase the value of $g$  and thereby increase the value of $V$ or $J_{1z}$ (see Table \ref{tab1}),
 the numerical calculation suffers from appreciable slowing down and  with our computational 
constraints we cannot study for large values of $g$ (i.e., $g\gtrsim2.25$).

\begin{figure}
\begin{center}
\includegraphics[width=3.0in,height=2.0in]
{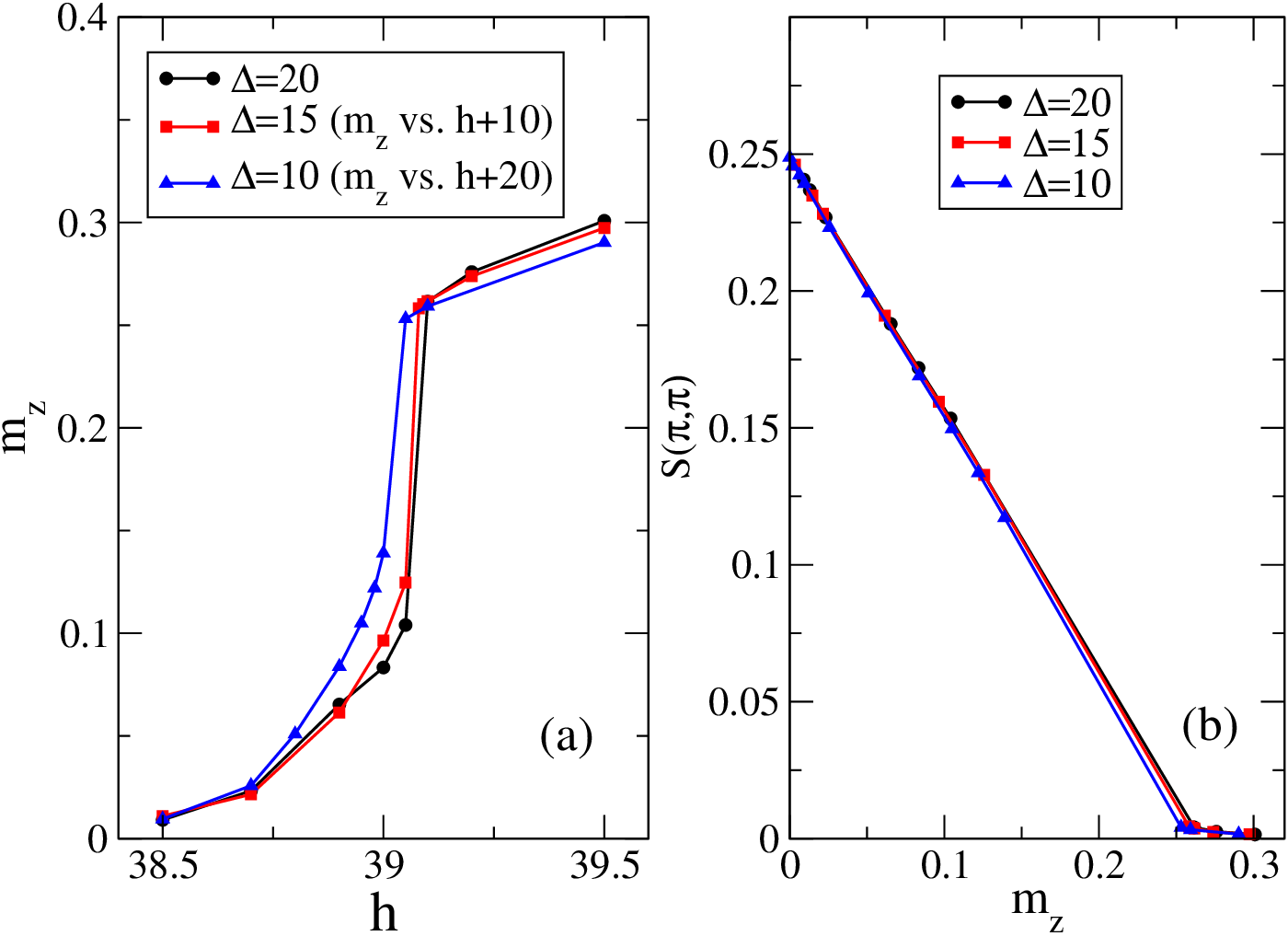}
\end{center}
\caption{(Color online) {\bf Determining cut-off for the anisotropy parameter
${\mathbf {\Delta = J_{1z}/J_{1xy}}}$ when ${\mathbf {J_{2xy}\neq0}}$.} Results of (a) $m_z$ vs h and (b) $S(\pi,\pi)$ vs $m_z$ for 
 a $16\times16$ lattice, for $t/\omega_0 =1$, and for $g=2.5$ but with $\Delta=10,~15$ and 20.}
\label{Figure 4}
\end{figure}

{
We can  set a cut-off for the anisotropy parameter $\Delta=J_{1z}/J_{1xy}$ above
 which the essential physics for the system does not change much; for values of $\Delta$ above the cut-off,
we expect that NN
occupancy is in effect projected out.} More precisely, we find that 
we can deal with large values of $g$ (i.e., $g \gtrsim  2.25$) by setting  $J_{1z}/J_{1xy}=\Delta_0=15$ 
and still get the correct behavior
 of the observables thereby saving computational time. 
Fig.~\ref{Figure 4} shows $m_z$ versus $h$ and $S(\pi,\pi)$ 
 against $m_z$ for
 $g=2.5$; results for different values of $\Delta$ are compared in order to 
decide on a cut-off value $\Delta=\Delta_0$. {In the large anisotropic limit,
 a change in the anisotropy parameter ${\rm d}\Delta$ can be shown to produce an additional
 effective field of $h_{\rm eff}=2{\rm d}\Delta$ (at the mean-field level). So in Fig. \ref{Figure 4}(a), 
we  shifted the $h$ scale accordingly in an attempt to make the $m_z$ plots coincide.}
The good agreement between the two cases for $\Delta=15$ and $\Delta = 20$ gives us the freedom to
 use a cut-off of $\Delta=\Delta_0=15$ at large values of $g$ (i.e., $g \gtrsim  2.25$).

\begin{figure}[b]
\vspace{.25 in}
\begin{center}
\includegraphics[
width=3.0in,height=1.8in]{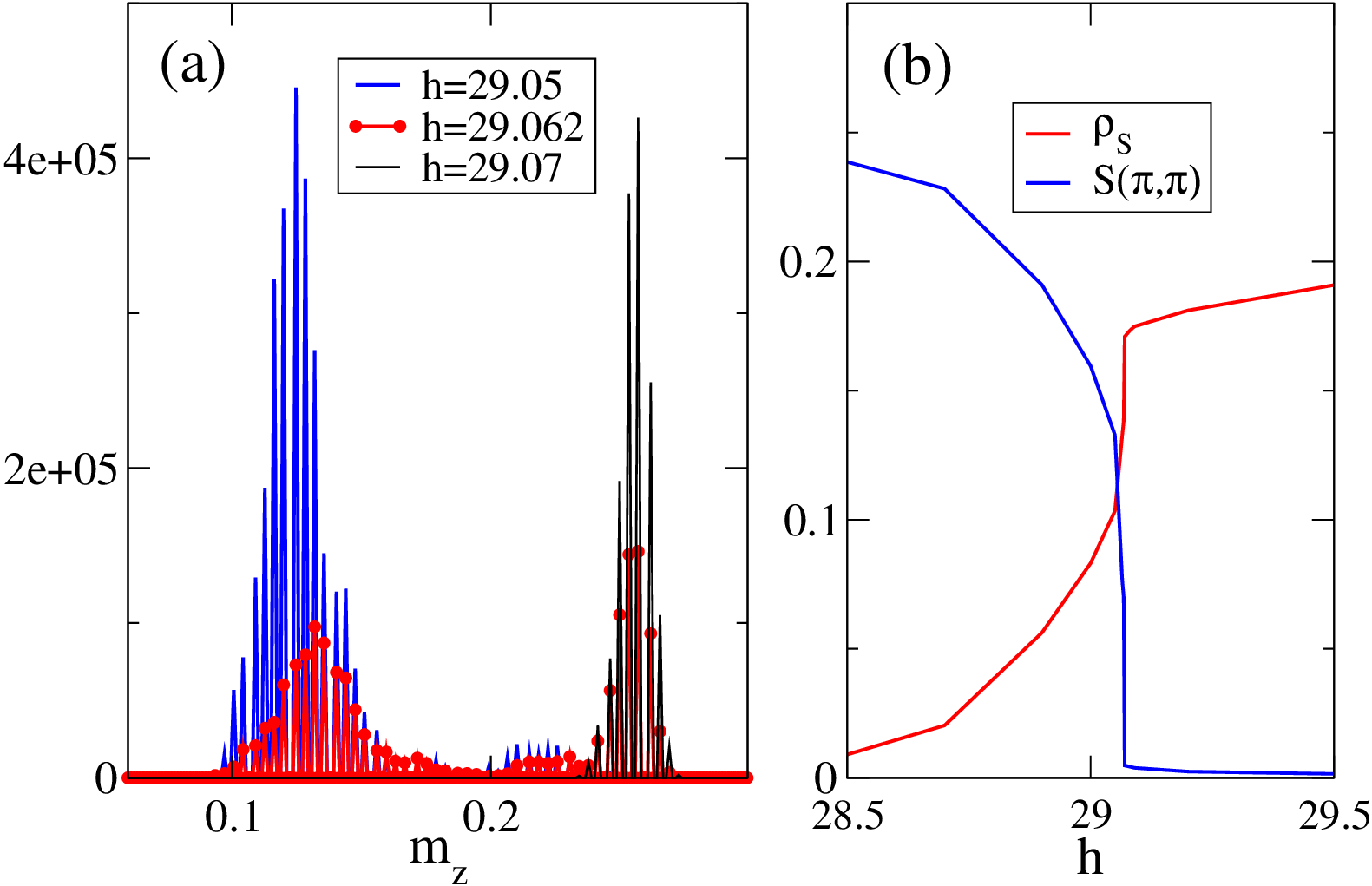}
\end{center}
\caption{(Color online) {\bf First-order transition shown by the effective BH Hamiltonian
when same-sublattice transport is inadequate.}
(a) Magnetization $m_z$ histograms in the  transition region showing phase separation
through a double-peaked structure at $h=29.062$; (b) evolution of $\rho_S$ and $S(\pi,\pi)$ vs $h$
during phase transition. 
Both plots are for $g=2.5$ and $t/\omega_0=1$.}
\label{Figure 5}
\end{figure}

\begin{figure}[htb]
\centering
\includegraphics[width=3.0in,height=3.6in]{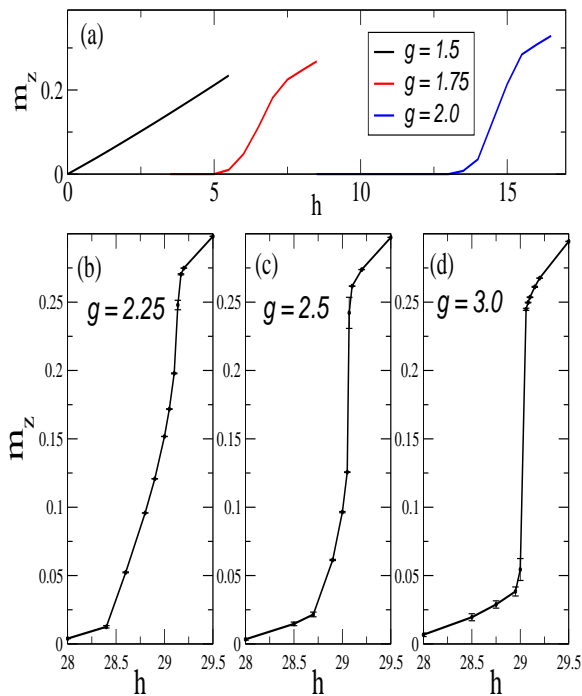}
\caption{(Color online) {\bf Evolution from continuous transition to discontinuous transition
as coupling strength increases in the  BH model.} Magnetization vs magnetic field in a $16\times16$ lattice for $t/\omega_0=1$ and for
different values of $g$.}
\label{Figure 6}
\end{figure}

With the above simplification, we do the SSE-QMC simulation for larger values of $g$
(such as $g \geq 2.25$) and observe  phase-separated phases; we explain this based on Fig.~\ref{Figure 5}
plotted at $g=2.5$. At values of magnetic field $h < 29.05$, a single peaked structure occurs. On increasing h, at $h=29.062$,
a double-peaked structure results showing simultaneous existence of two phases (with magnetizations centered at $m_z \approx .12$ and $m_z \approx .25$).
A further small increase to $h=29.07$, leads to again a single peak (centered around $m_z\approx 0.26$) signalling that 
a discontinuous phase transition has occurred. 
Thus a phase separated state at $g=2.5$ is clearly captured in Fig.~\ref{Figure 5}(a).
In Fig.~\ref{Figure 5}(b), we show the evolution of the SF density $\rho_S$ and $S(\pi,\pi)$ as $h$ is varied
and capture the first-order transition at $h \approx 29.06$.
 
\begin{table}
\caption{Autocorrelation times at $g=2.5$ and $t/\omega_0 =1$.}
\parbox{.48\linewidth}{
\centering
\begin{tabular}
  {|p{35pt}| p{35.00pt}| p{35.00pt}|}   \hline
 {$\Delta=10$}
 & { $\epsilon=1.0$} & { $\epsilon=2.5$}
  \\ \hline\hline ${\rm h}=18.0$ & 250289 & 96613 \\ \hline ${\rm h}=18.5$
& 658170 &
174995 \\ \hline ${\rm h}=19.0$ & 208295 &  9835 \\ \hline ${\rm h}=19.5$ &
960 &  113 \\
\hline
\end{tabular}
}
\hfill
\parbox{.48\linewidth}{
\centering
\begin{tabular}  {|p{35pt}| p{35.00pt}|p{35.00pt}|}   \hline
 {$\Delta=15$}
 & { $\epsilon=4.0$}& { $\epsilon=6.0$}
  \\ \hline\hline ${\rm h}=28.0$ & 238406 & 82608 \\ \hline ${\rm h}=28.5$
& 856138 & 666353 \\ \hline ${\rm h}=29.0$ & 161336  & 8646 \\ \hline
${\rm h}=29.1$ & 4847 & 4778 \\
\hline
\end{tabular}
}
\label{autocor}
\end{table}

The results for the magnetization versus magnetic field in the transition region for values of  $g=2.25,~2.5$ and $3.0$ 
are shown in Figs.~\ref{Figure 6}(b), (c), and (d). There is a  gradual increase in the quantum of the magnetization jump
as the coupling $g$ is increased. 
 We should also mention here that, particularly in the calculations 
involving large $\Delta$, we wanted to  ensure that the numerical data in adjacent bins are not correlated.
To this end, we compute the autocorrelation time  $\tau_{int}$  defined as \cite{sse1}
\begin{eqnarray*}
\tau_{int}[m_z]=\frac{1}{2}+\sum_{t=1}^\infty A_{m_z}(t) ,
\end{eqnarray*}
where 
\begin{align}
A_{m_z}(t)=\frac{<m_z(i+t)m_z(i)>-<m_z(i)>^2}{<m_z(i)^2>-<m_z(i)>^2} ,
\end{align}
with $i$ and $t$ being Monte Carlo times defined in units of Monte
Carlo sweeps (MCS).

Typically, we see that for moderately high values of $\Delta$, $\epsilon=\Delta/4$ can
 restrict $\tau_{int}$ from attaining very large values. Thus, for $\Delta\le10$, we 
use a large bin size (i.e., 15,00,000 MCS) in our simulations and keep  $\tau_{int}$ 
 well within the bin size in order to produce meaningful results from our  simulation.
However, for larger $\Delta$, even choosing $\epsilon=\Delta/4$ cannot keep
 autocorrelation times sufficiently smaller than such large bin sizes.
 So, in those cases, we use larger $\epsilon$ values (i.e., $\epsilon$=6 and 8 for $\Delta$=15 and 20, respectively)
 and even larger bin sizes (i.e., 22,00,000 MCS).
The values of the autocorrelation times, for $g=2.5$ and at various fields $h$ close to the transition, 
are shown in Table~\ref{autocor}. It should also be noted that we
 cannot  take $\epsilon$ too large, as a calculation with $\epsilon$  
larger than $\Delta$ does not produce meaningful results.

Lastly, we mention that our phase diagrams, both for the XXZ model and our extension of it, are similar (though not identical)
 to what were obtained using Lanczos method in Ref. \onlinecite{sanjoy}.

\section{Summary}
In this work, we studied  the  effective Hamiltonian
of a Bose-Holstein model
 using the SSE-QMC technique and obtained the ground state phase diagram.
We found that supersolidity is realized at intermediate couplings; whereas, at large couplings
the system phase separates because the double-hopping terms (that produce transport in the same sublattice)
 are not dominant.

Our results on a large $16\times16$ lattice {represent well the system behavior 
in the thermodynamic limit, as demonstrated in Fig. \ref{Figure 1} using different finite-size calculations.} 
Our results are only qualitatively similar to those obtained
 earlier  using modified Lanczos technique on a much smaller $4\times4$ cluster \cite{sanjoy}.

We overcame computational difficulties for large repulsive interactions
by devising a cutoff repulsive strength; above the cutoff, the system
properties (as shown in Fig. \ref{Figure 4})
 become essentially
independent of the strength of repulsion
(because repulsive interactions  project out nearest-neighbor occupation). 
 To 
mimic the results of statistically independent configurations,
we considered
 sufficiently large $\epsilon$ values and kept the auto-correlations within acceptable limits.

Our work is an exercise in
 SSE-QMC study of a simple but  important model which 
has significance  in various 
fields. 
We hope that the present results will stimulate further investigations in
allied areas such as frustrated quantum magnets in various geometries and at various magnetizations;
coherence dynamics of excitons/spins
in the presence of phonon environments (pertinent to quantum computation and artificial light harvesting);
 dimer formation and dimer correlations in Hubbard-Holstein
model \cite{murakami}, HCBs coupled to multimode phonons \cite{zoller}, etc.

\appendix

\section{SSE bond operators}

In SSE-QMC study of Heisenberg spin systems, the Hamiltonian is
written as a bond Hamiltonian. Particularly, in our case we write
$H=-\sum_{i=1}^3\sum_b H_{b_i}$ where $b_1,~b_2,~\&~b_3$ denote the NN,
NNN, and NNNN bonds in our spin model, respectively. Each of such
$H_{b_i}$ consists of the diagonal ($H_{1,{b_i}}$) and the off-diagonal $(H_{2,{b_i}}$) parts and is given as 
$H_{b_i}=H_{1,{b_i}}+H_{2,{b_i}}$ with 
expressions
\begin{align}
H_{1,b_i}&=C_i-J_{iz}S^z_{i(b_i)}S^z_{j(b_i)}+h_b[S^z_{i(b_i)}+S^z_{j(b_i)}]\nonumber\\
H_{2,b_i}&=-\frac{J_{ixy}}{2}[S^+_{i(b_i)}S^-_{j(b_i)}+{\rm H.c.}] ,
\end{align}
where $J_{2z}=J_{3z}=0$, $C_i\equiv J_{iz}/4+h_b+\epsilon J_{1xy}$, $\epsilon\geq 0$, 
and $h_b\equiv hJ_{1xy}/z$ with the coordination number $z=12$.
In our model, a two-spin matrix element of any of these
operators can never become negative.

\vspace{-0.4cm}

\section*{Acknowledgements}
The authors thank Pinaki Sengupta and Keola Wierschem for useful discussions on the SSE technique and
its implementation for our HCB system. We also thank Amrita Ghosh for 
helping with the calculations and for cross-checking.




\section*{}



\setcounter{equation}{0}
\setcounter{figure}{0}
\renewcommand{\theequation}{S\arabic{equation}}
\renewcommand{\thefigure}{S\arabic{figure}}
\noindent {{\bf Supplementary Material for \\
\\ 
\indent ``Coexistence of long-range orders in a Bose-Holstein model''}\\
\\
\indent Satyaki Kar and Sudhakar Yarlagadda}
\begin{center}
{\bf Comparison between ED and SSE results}
\end{center}

\begin{figure}[htb]
\vspace{.25 in}
\centerline{\includegraphics[
width=2.8in,height=2.in]{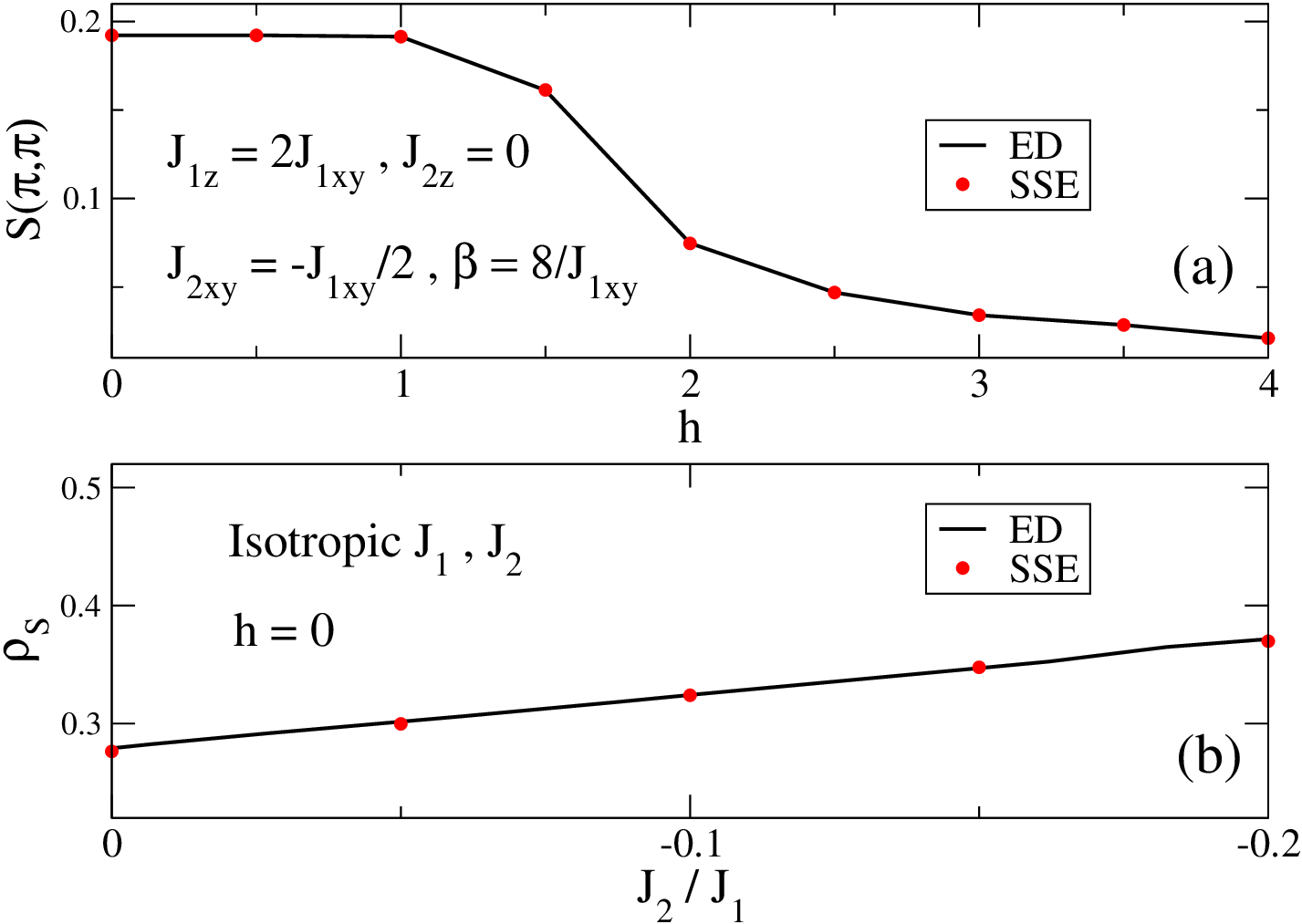}}
\caption{(Color online) Comparison of SSE with ED results on $4\times 4$ clusters for a 2D $J_1-J_2$ model:
 (a) S($\pi,\pi$) vs h (ED results are obtained using LAPACK),
(b) $\rho_S$ vs $J_2/J_1$ (ED results are taken from Refs. \onlinecite{einarsson,runge}). }
\label{Figure 7}
\end{figure}

We have benchmarked our QMC calculations by comparing our calculated values with those obtained by exact-diagonalization (ED) methods.
 We consider both the XXZ model and its simple extension, namely, the  $J_1-J_2$ model. We find that the energy, magnetization, structure factor $S(\pi,\pi)$ 
and SF  density $\rho_S$ of our SSE calculations match quite well with those from the ED results. 
The comparisons of the calculated $S(\pi,\pi)$ and $\rho_S$
for the $J_1-J_2$ model 
 are shown in Fig.~\ref{Figure 7}(a)-(b).

Our SSE results also compare well with various world-line Monte Carlo
 results\cite{scalettar1,scalettar2,scal3} and also the SSE QMC results\cite{chen,wessel} available in the literature.

\end{document}